\documentclass{sig-alternate-2013}

\usepackage{graphicx}
\usepackage[font=bf]{caption}
\usepackage{subcaption}

\usepackage[USenglish]{babel}
\usepackage{microtype}

\usepackage{csquotes}
\MakeOuterQuote{"}

\usepackage{tabulary}
\usepackage{booktabs}
\usepackage{cellspace}
\usepackage{balance}

\def\sharedaffiliation{
\end{tabular}
\begin{tabular}{c}
}

\permission{\copyright 2017 International World Wide Web Conference Committee  (IW3C2), \\ published under Creative Commons CC BY 4.0 License.}
\conferenceinfo{WWW 2017 Companion,}{April 3--7, 2017, Perth, Australia.}
\copyrightetc{ACM \the\acmcopyr}
\crdata{978-1-4503-4913-7/17/04. \\
http://dx.doi.org/10.1145/3041021.3054202 \\
\includegraphics{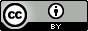}}

\begin{document}


\title{Related Pins at Pinterest: \\
The Evolution of a Real-World Recommender System}
\author{
    David C. Liu, 
    Stephanie Rogers, 
    Raymond Shiau, 
    Dmitry Kislyuk, \\
    Kevin C. Ma, 
    Zhigang Zhong, 
    Jenny Liu, 
    Yushi Jing
    \sharedaffiliation
    \affaddr{Pinterest, Inc.} \\
    \affaddr{San Francisco, CA} \\
    \affaddr{\{dliu,srogers,rshiau,dkislyuk,kevinma,az,jennyliu,jing\}@pinterest.com}
}

\maketitle

\begin{abstract}
Related Pins is the Web-scale recommender system that powers over 40\% of user engagement on Pinterest.
This paper is a longitudinal study of three years of its development, exploring the evolution of the system and its components from prototypes to present state.
Each component was originally built with many constraints on engineering effort and computational resources, so we prioritized the simplest and highest-leverage solutions.
We show how organic growth led to a complex system and how we managed this complexity.
Many challenges arose while building this system, such as avoiding feedback loops, evaluating performance, activating content, and eliminating legacy heuristics.
Finally, we offer suggestions for tackling these challenges when engineering Web-scale recommender systems.
\end{abstract}



\section{Introduction}

Much literature has been published on advanced recommendation systems as well as their real-world applications.
However, it is usually not possible to build state-of-the-art recommender systems directly.
The initial product must be built with a small engineering team, limited computational resources, and no training data until the recommender is bootstrapped.
Industry recommenders often deal with Web-scale data comprising billions of items.
The content is often poorly labeled and training data is noisy because it is collected through implicit user feedback~\cite{hu2008implicit}.
As a result, many practitioners choose to use ad-hoc heuristics and make trade-offs when building the initial system.
However, trying to grow the system can quickly complicate it, making it difficult to reason about further changes.

\begin{figure}[htbp]
    \centering
    \begin{subfigure}[t]{\columnwidth}
        \includegraphics[width=0.4\columnwidth]{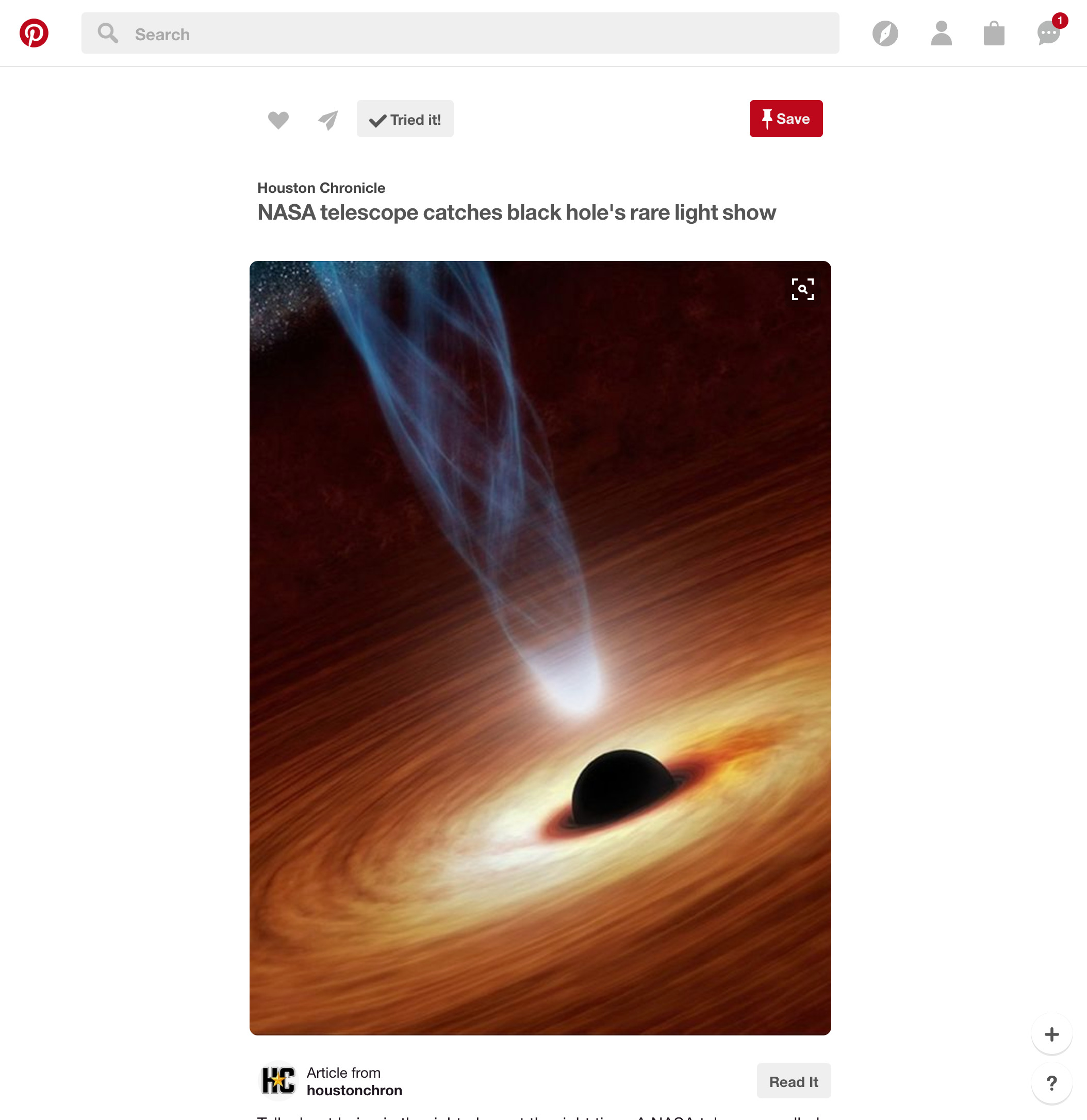}
        \includegraphics[width=0.52\columnwidth]{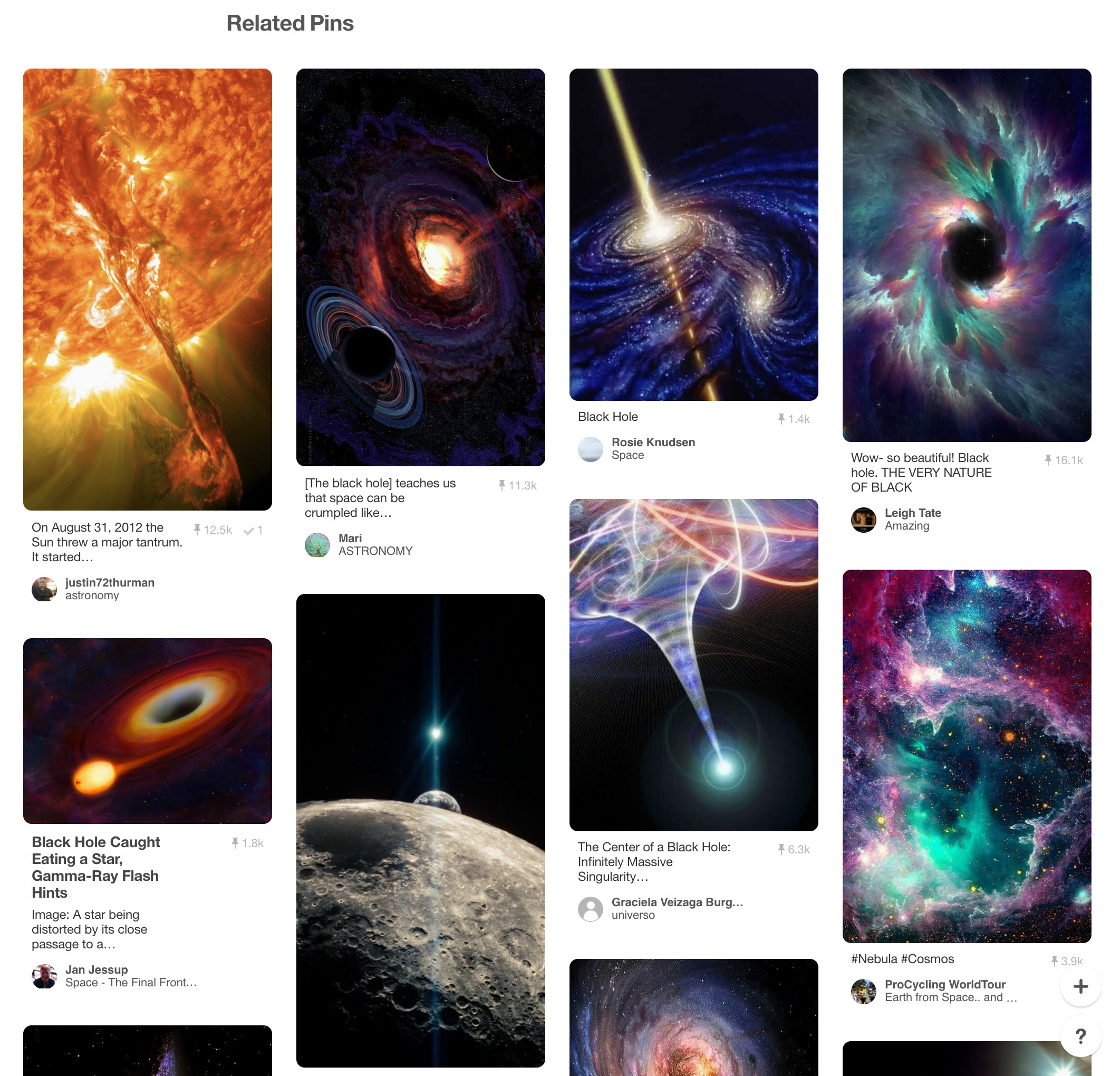}
        \caption{Related Pins on the pin closeup view.}
        \label{fig:surfaces/pin-closeup}
    \end{subfigure}
    \begin{subfigure}[t]{0.9\columnwidth}
        \includegraphics[width=\columnwidth]{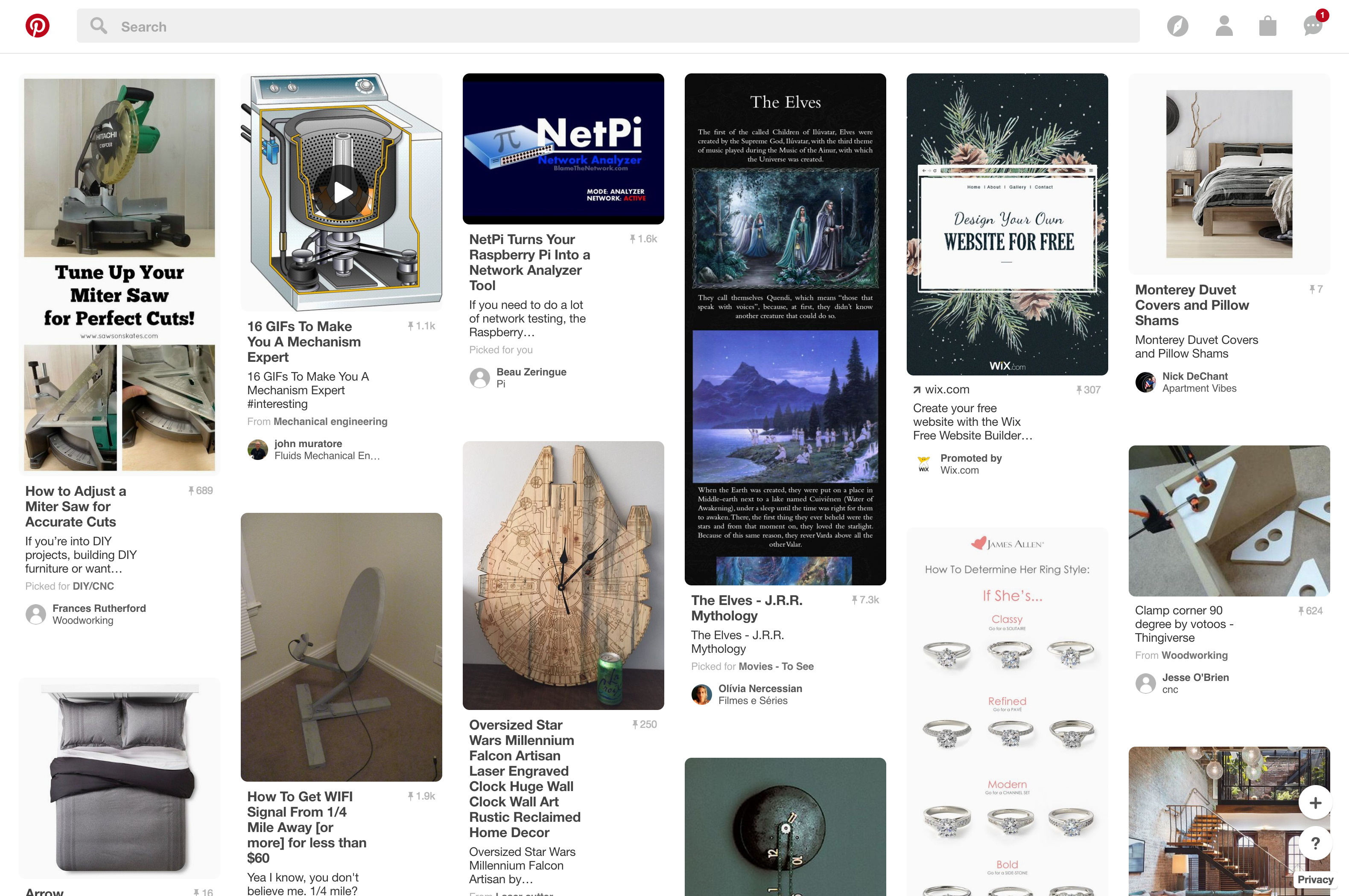}
        \caption{Some pins on Home Feed are from Related Pins of recent user activity.}
        \label{fig:surfaces/home-feed}
    \end{subfigure}
    \begin{subfigure}[t]{0.9\columnwidth}
        \includegraphics[width=\columnwidth]{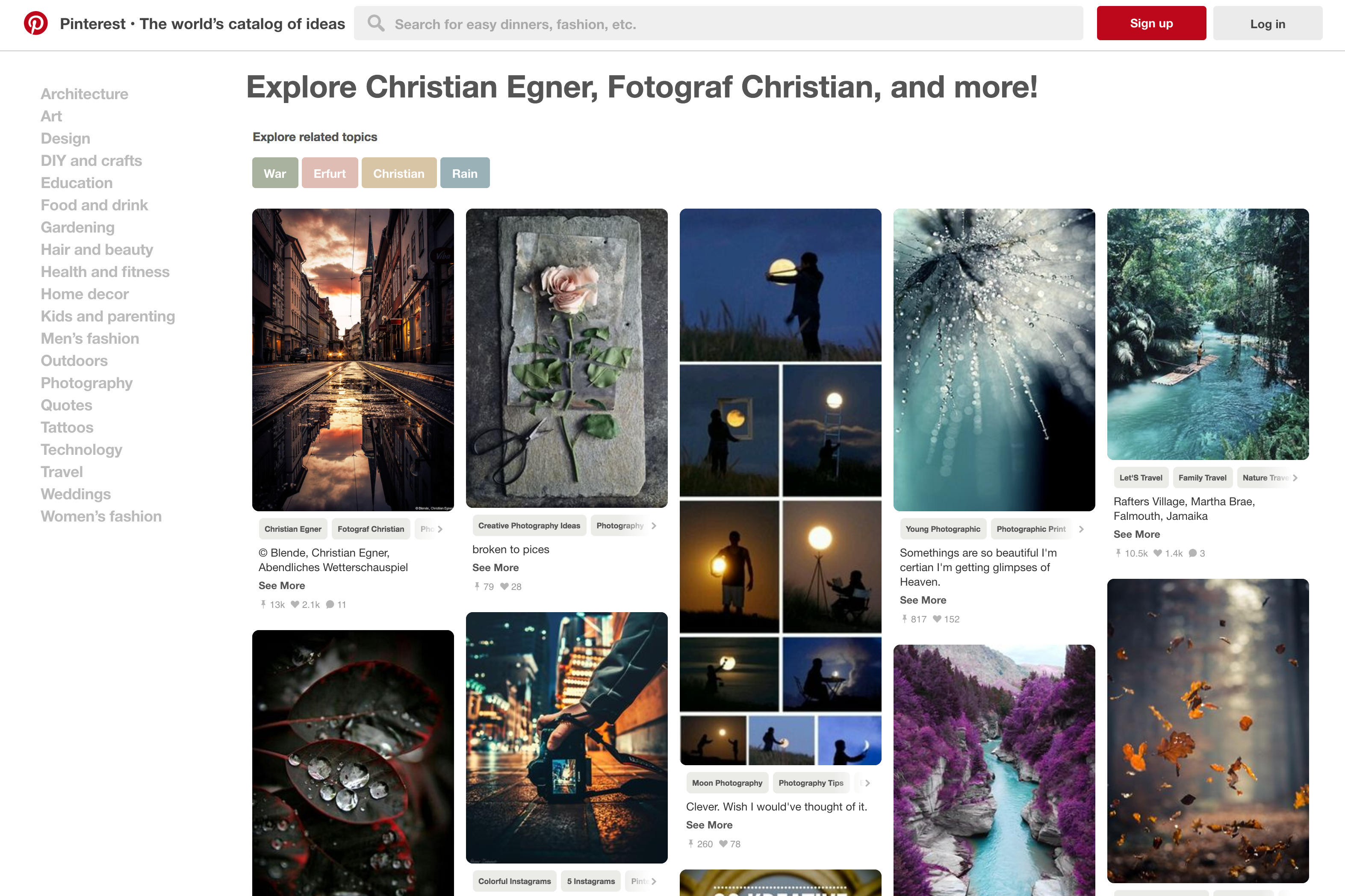}
        \caption{The unauthenticated pin landing page showcases Related Pins to entice visitors to sign up.}
        \label{fig:surfaces/unauth-pin-page}
    \end{subfigure}
    \caption{Some major product surfaces featuring Related Pins.}
    \label{fig:surfaces}
\end{figure}

At Pinterest, we had the unique opportunity to observe these problems over a time horizon of three years, in the context of Related Pins.
The initial version of Related Pins, launched in 2013, was one of the first forays into recommender systems at Pinterest.
Though successful in improving content discovery, Related Pins initially received minimal engineering attention.
In 2014, about 10\% of pins saved across Pinterest were discovered through Related Pins.
In 2015, a small team began iterating and further developing Related Pins.
It now drives over 40\% of all saves and impressions through multiple product surfaces, and it is one of the primary discovery mechanisms on Pinterest.
This paper explores the challenges of real-world recommender systems through a longitudinal study of Related Pins.
In describing the gradual evolution of our system we present solutions for these challenges, rationales for our trade-offs, and key insights learned.

Real-world recommender systems have been described for music suggestion~\cite{cai2007scalable}, image search~\cite{jing2008pagerank}, video discovery on YouTube~\cite{baluja2008video}\cite{deepyoutube}\cite{Davidson:2010} and movies on Netflix~\cite{Gomez-Uribe}. 
Many of these papers describe final systems; however, they do not describe how one might build the system incrementally.
Many challenges facing real-world systems are described in \cite{sculleyhidden}.
We provide concrete examples of how these challenges arise in Related Pins and propose unique solutions.

For Related Pins, we prioritize shipping the simplest, highest-leverage products first, in order to reach incremental milestones and demonstrate viability.
Our original recommendation algorithm consisted of a simple candidate generator with many heuristic rules.
Though it was built in just three weeks, it capitalized on the strong signal present in user-curated boards.
We continued adding more candidate sources as we identified gaps in the coverage and recall (Section~\ref{sec:candidates}).
Over time, we introduced a memorization layer to boost popular results.
\textit{Memboost} is lightweight, both in engineering complexity and computational intensity, yet significantly leverages a vast amount of user feedback.
We had to account for position bias and deal with complexity in the form of feedback loops, but found the benefits worth the cost (Section~\ref{sec:memboost}).
Next, we added a machine-learned ranking component because we thought it would have the most potential for impact.
We started with a basic linear model with just nine features.
As we found shortcomings of the model and our training methodology, we began to experiment with more advanced approaches (Section~\ref{sec:ranking}).





\section{Related Pins System Overview}\label{sec:overview}

Pinterest is a visual discovery tool for saving and discovering content. 
Users save content they find on the Web as \textit{pins} and create collections of these pins on boards.
Related Pins leverages this human-curated content to provide personalized recommendations of pins based on a given \textit{query pin}. 
It is most prominent on the \textit{pin closeup} view shown in Figure~\ref{fig:surfaces}.
Related Pins recommendations are also incorporated into several other parts of Pinterest, including the home feed, pin pages for unauthenticated visitors, the "instant ideas" button for related ideas~\cite{sharp2017instantideas}, emails, notifications, search results, and the "Explore" tab.

User engagement on Pinterest is defined by the following actions.
A user \textit{closeups} on a pin by clicking to see more details about the pin.
The user can then \textit{click} to visit the associated Web link; if they remain off-site for an extended period of time, it is considered a \textit{long click}.
Finally, the user can \textit{save} pins onto their own boards.
We are interested in driving "Related Pins Save Propensity," which is defined as the number of users who have saved a Related Pins recommended pin divided by the number of users who have seen a Related Pins recommended pin.

In the Pinterest data model, each pin is an instance of an image (uniquely identified by an image signature) with a link and description.
Although each pin is on a single board, the same image can be used in many pins across different boards: when a pin is saved to a new board, a copy of the pin is created.
Pin information is typically aggregated on the image signature level, providing richer metadata than individual pin instances.
For convenience, future references to "query pin" and "result pin" actually refer to the aggregation of pins with the same image signature.

\begin{figure}[htbp]
    \centering
    \includegraphics[width=\columnwidth]{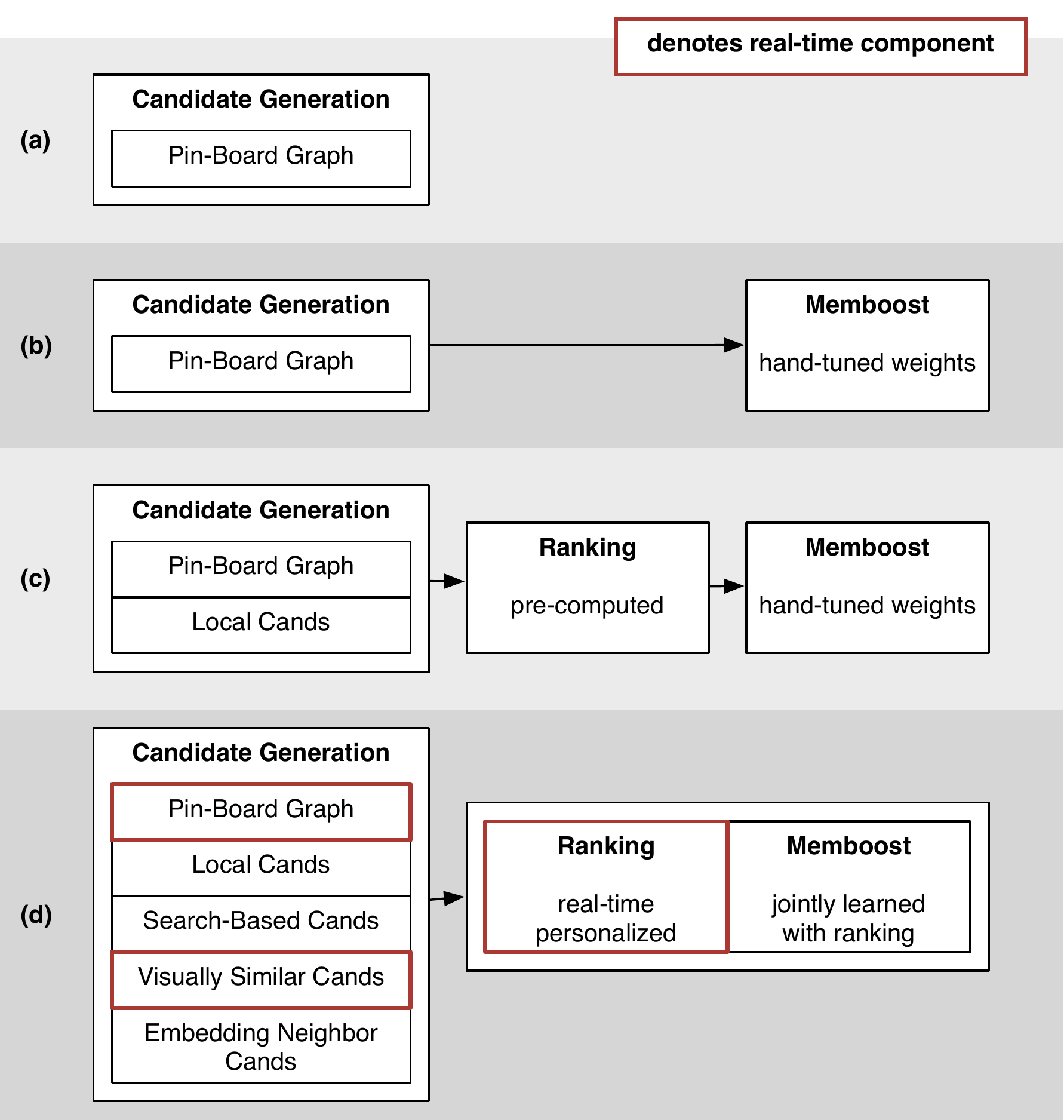}
    \caption{Snapshots of our system architecture over time.}
    \label{fig:snapshots}
\end{figure}

The Related Pins system comprises three major components summarized below.
The components were introduced to the system over time, and they have each evolved dramatically in their own right.
Figure~\ref{fig:snapshots} shows various snapshots of our architecture illustrating the evolution of both the overall system as well as each of the three components. 
Subsequent sections of the paper explore their development in more detail.

\textbf{Candidate generation.} 
We first narrow the candidate set---the set of pins eligible for Related Pin recommendations---from billions to roughly 1,000 pins that are likely related to the query pin. We have developed and iterated on several different candidate generators to do this.

\textbf{Memboost.}
A portion of our system memorizes past engagement on specific query and result pairs. 
We describe how we account for position bias when using historical data, by using a variant of clicks over expected clicks~\cite{zhang2007coec}. 
Introducing memorization increases system complexity with feedback loops, but significantly boosts engagement.

\textbf{Ranking.}
A machine-learned ranking model is applied to the pins, ordering them to maximize our target engagement metric of Save Propensity.
It uses a combination of features based on the query and candidate pins, user profile, session context, and Memboost signals. 
We apply learning-to-rank techniques, training the system with past user engagement.


\section{Evolution of Candidates}\label{sec:candidates}

The original Related Pins system consisted of just one form of candidate generation: leveraging the graph of pin and boards by extracting frequently co-occurring pins.
These candidates were shown directly to the user as recommendations (Figure~\ref{fig:snapshots}a).
Later, when we introduced Memboost and machine-learned ranking, the problem of candidate generation shifted from precision to recall: generating a diverse set of pins relevant to the query pin. 
This led us to add new candidate sources as we identified gaps in coverage and recall (Figure~\ref{fig:snapshots}d).

\subsection{Board co-occurrence}
Our primary candidate generator continues to be based on the user-curated graph of boards and pins, but we've changed the method over time to produce more relevant results and to cover more query pins.

\begin{figure*}[htbp]
    \centering
    \includegraphics[width=\textwidth]{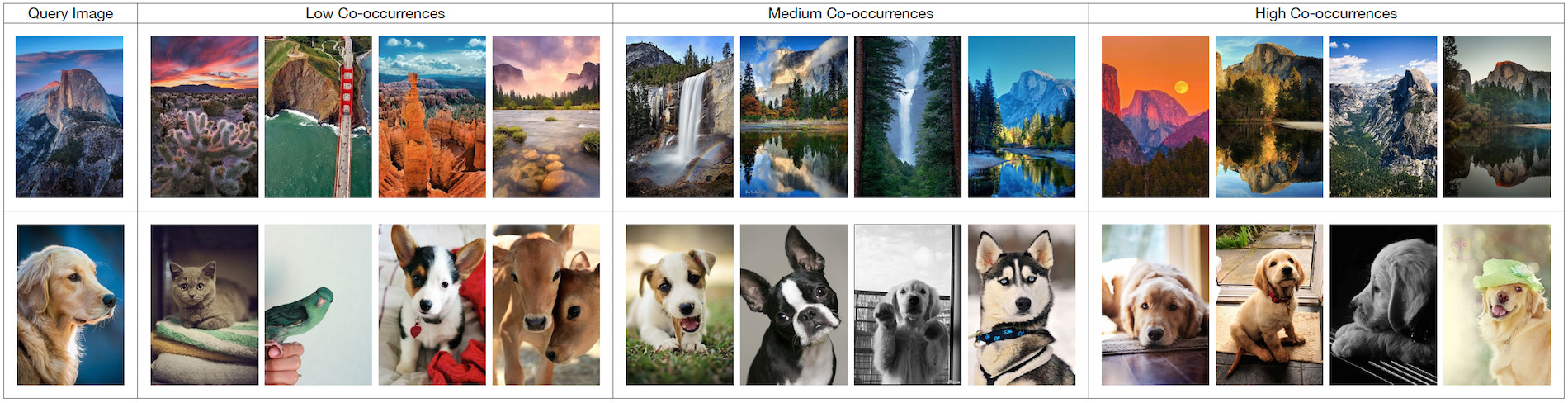}
    \caption{Examples of candidates with low, medium and high board co-occurrences with the query image. \textit{Top example:} low: travel destinations, medium: Yosemite viewpoints, high: Half Dome. \textit{Bottom example:} low: animals, medium: dogs, high: golden retrievers ~\cite{kislyuk2015human, yliu2016power}.}
    \label{fig:occur_examples}
\end{figure*}

\subsubsection{Heuristic Candidates} 
The original Related Pins were computed in an offline Hadoop Map/Reduce job: we mapped over the set of boards and output pairs of pins that occurred on the same board.
There are too many pairs of possible pins, so pairs are randomly sampled to produce approximately the same number of candidates per query pin.
We further added a heuristic relevance score, based on rough text and category matching.
The score was hand-tuned by inspecting example results.

We chose this method for its simplicity. 
Because of limited engineering resources, it was built by two engineers in just three weeks.
Additionally, because the human-curated board graph was already a very strong signal~\cite{yliu2016power}, it proved to be a rather effective method.

\subsubsection{Online Random Walk}

We found that the relevance of candidates qualitatively increases with higher board co-occurrence, as shown in Figure~\ref{fig:occur_examples}.
However, the original method was primarily based on the heuristic score; it did not attempt to maximize board co-occurrence.
We also noted that rare pins, pins occurring on only a few boards, did not have many candidates.
To address these limitations we moved to generating candidates at serving time through an online traversal of the board-to-pin graph.

Candidates are now generated by a random walk service called Pixie \cite{pixie}.
A full description is outside the scope of this paper, but broadly, Pixie loads the bipartite graph of pins and boards into a single machine with large memory capacity.
The edges of the graph represent individual instances of a pin on a board.
The graph is pruned according to some heuristic rules to remove high-degree nodes and low-relevance pins from boards.
Pixie conducts many random walks (on the order of 100,000 steps) on this graph starting from the query pin, with a reset probability at each step of the walk, and aggregates pin visit counts (similar to \cite{gupta2013wtf}).
This effectively computes Personalized PageRank on the graph seeded with the query pin.

This system is much more effective at leveraging board co-occurrence, since highly connected pins are more likely to be visited by the random walk.
It also increases candidate coverage for rare pins, since it can retrieve pins that are several hops away from the query.

\subsection{Session Co-occurrence}
Board co-occurrence offers good recall when generating candidates, but the rigid grouping of boards suffers inherent disadvantages.
Boards are often too broad, so any given pair of pins on a board may only be tangentially related.
This is especially true of long-lived boards, as the topic of a board will drift with the user's interest.
Boards may also be too narrow: for example, a whiskey and a cocktail made with that whiskey might be pinned in close succession to different boards.
Both these shortcomings can be addressed by incorporating the temporal dimension of user behavior: pins saved during the same session are typically related in some way.
We built an additional candidate source called Pin2Vec~\cite{ma2017pin2vec} to harness these session co-occurrence signals.

\begin{figure}[htbp]
    \centering
    \includegraphics[width=\columnwidth]{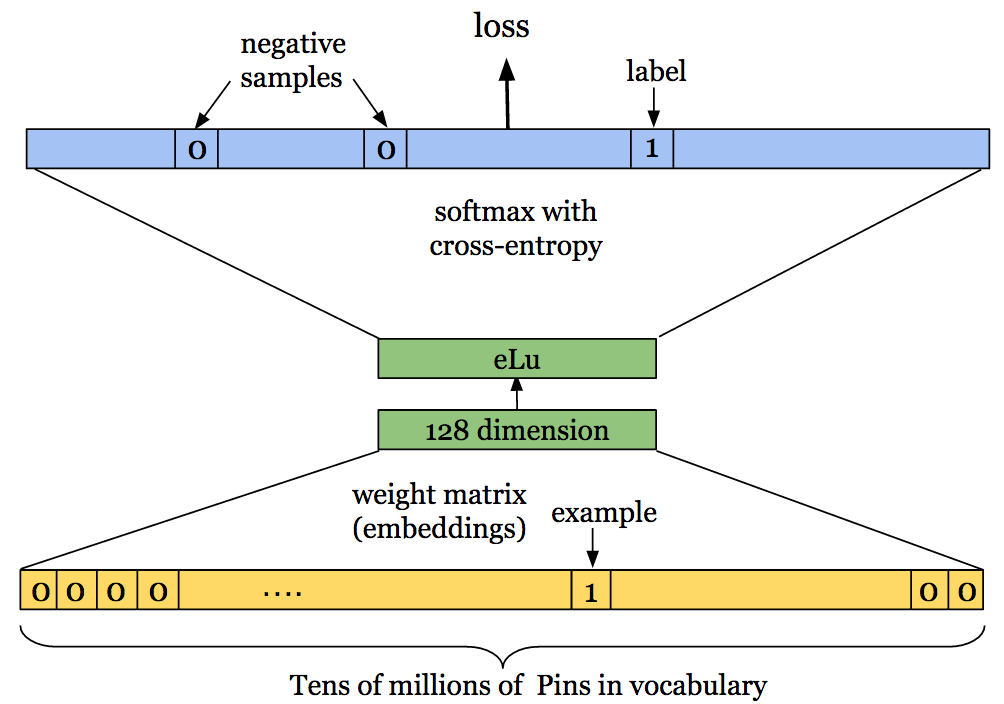}
    \caption{Neural network architecture of Pin2Vec.}
    \label{fig:pin2vec}
\end{figure}

Pin2Vec is a learned embedding of the $N$ most popular (head) pins in a $d$-dimensional space, with the goal of minimizing the distance between pins that are saved in the same session.
The architecture of the neural network is similar to word2vec~\cite{mikolov2013word2vec}.
The learning problem is formulated as a $N$-way classification, where the input and output are both one of $N$ pins (Figure~\ref{fig:pin2vec}).

To produce training data, we consider pins that are saved by the same user within a certain time window to be related.
Each training example is a pair of such pins.
Given one of the pins as input, an embedding matrix maps pin IDs to vectors in $\mathbb R^d$, and a softmax layer is used to map the embedding back into a predicted output pin ID.
The other pin in the pair is given as the expected output, and we train the embedding by minimizing the cross-entropy loss of the network.
Negative examples are sampled to make the optimization tractable.
The model is built and trained using TensorFlow~\cite{tensorflow}, and the result is a $d$-dimensional embedding for each of the $N$ pins.
At serving time, when the user queries one of the $N$ pins, we generate candidate pins by looking up its nearest neighbors in the embedding space.
We found that introducing these session-based candidates in conjunction with board-based candidates led to a large increase in relevance when one of the $N$ pins is used as a query. Conceptually, it captures a large amount of user behavior in a compact vector representation.


\subsection{Supplemental Candidates}
In parallel with the above progress, we started developing new candidate generation techniques for two reasons.
First, we wanted to address the \textit{cold start} problem: rare pins do not have a lot of candidates because they do not appear on many boards. 
Second, after we added ranking (Section~\ref{sec:ranking}), we wanted to expand our candidate sets in the cases where diversity of results would lead to more engagement.
For these reasons, we started to leverage other Pinterest discovery technologies.

\textbf{Search-based candidates.}
We generate candidates by leveraging Pinterest's text-based search, using the query pin's annotations (words from the web link or description) as query tokens.
Each popular search query is backed by a precomputed set of pins from Pinterest Search. 
These search-based candidates tend to be less specifically relevant than those generated from board co-occurrence, but offer a nice trade-off from an exploration perspective: they generate a more diverse set of pins that are still somewhat related.

\textbf{Visually similar candidates.}
We have two visual candidate sources, described further in \cite{jing2015visual} and \cite{zhai2017visualDisco}.
If the query image is a near-duplicate, then we add the Related Pins recommendations for the duplicate image to the results.
If no near-duplicate is identified, then we use the Visual Search backend to return visually similar images, based on a nearest-neighbor lookup of the query's visual embedding vector.

\subsection{Segmented Candidates}\label{sec:candidates/segmented}
Finally, we wanted to address the \textit{content activation} problem: rare pins do not show up as candidates because they do not appear on many boards.

When Pinterest began focusing on internationalization, we wanted to show international users more results in their own language.
The majority of content was English and from the United States.
Although local content did exist, it was not very popular nor connected to popular pins, and thus was not generated by the other candidate sources.
To solve this, we generate additional candidate sets segmented by locale for many of the above generation techniques.
For example, for board co-occurrence, we filter the input set for the board-pin graph to only include pins of that given locale.
This methodology could be extended to other dimensions with content activation issues, too, such as gender-specific content or fresh content.


\section{Evolution of Memboost}\label{sec:memboost}

Initial versions of Related Pins already received a high amount of engagement. 
As a first step toward learning from our massive engagement logs, we built \textit{Memboost} to memorize the best result pins for each query.
We chose to implement it before attempting full-fledged learning, because it was much more lightweight and we intuitively believed it would be effective.

\begin{figure}[htbp]
    \centering
    \includegraphics[width=\columnwidth]{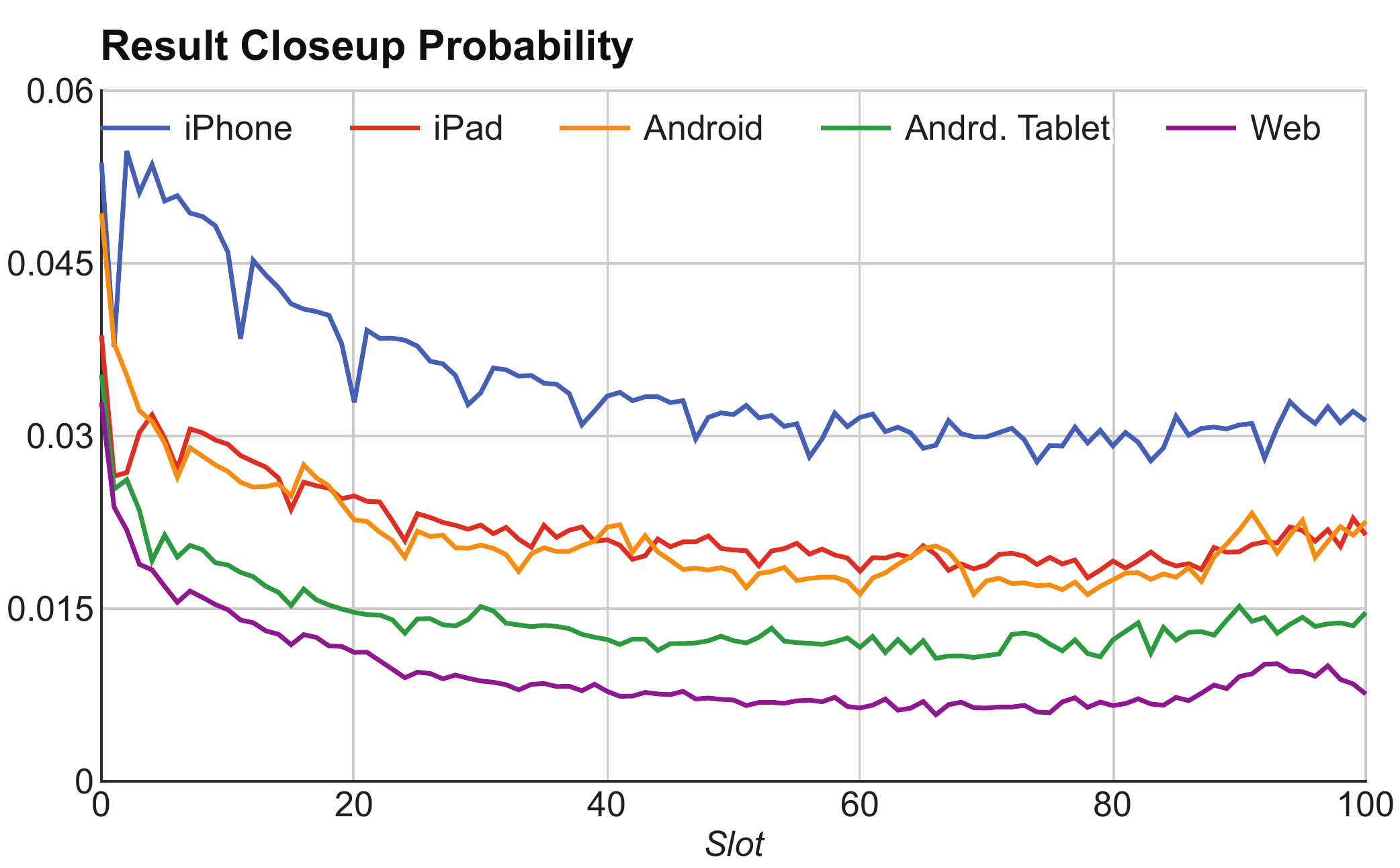}
    \caption{Global click-through rate of Related Pins by rank and platform.}
    \label{fig:ctr_prior}
\end{figure}

We initially wanted to simply incorporate the historical click-through rate of each result.
However, log data is subject to a strong \textit{position bias}: items shown in earlier positions are more likely to be clicked on.
Figure~\ref{fig:ctr_prior} illustrates this bias in the global click-through rate of each rank on each platform.
To account for this, we instead chose to compute \textit{clicks over expected clicks} (COEC) \cite{zhang2007coec}.
Let $\mathrm{clicks}(q, r)$ be the total number of clicks received by result pin $r$ on query pin $q$, and let $i_{p,k}(q, r)$ be the number of impressions it received on platform $p$ and rank $k$.
Each impression contributes a certain fractional number of \textit{expected clicks}, based on the global prior $\mathrm{clickrate}_{p,k}$ for that rank and platform.
The number of expected clicks for each result is  $\mathrm{Eclicks}(q,r) = \sum_p \sum_k i_{p,k}(q, r)\cdot \mathrm{clickrate}_{p,k}$.

We extended these definitions to other engagement actions, weighting the actions by $\beta_1, \dots, \beta_4$ as follows.
\begin{multline*}
\mathrm{actions}(q,r) =
    \beta_1 \cdot \mathrm{clicks}(q,r) +
    \beta_2 \cdot \mathrm{longclicks}(q,r) \\ +
    \beta_3 \cdot \mathrm{closeups}(q,r) +
    \beta_4 \cdot \mathrm{saves}(q,r)
\end{multline*}
\begin{multline*}
\mathrm{Eactions}(q,r) =
    \beta_1 \cdot \mathrm{Eclicks}(q,r) +
    \beta_2 \cdot \mathrm{Elongclicks}(q,r) \\ +
    \beta_3 \cdot \mathrm{Ecloseups}(q,r) +
    \beta_4 \cdot \mathrm{Esaves}(q,r)
\end{multline*}

Now, $\mathrm{actions}(q,r) / \mathrm{Eactions}(q,r)$ is similar to COEC generalized to all engagement actions.
To get a zero-centered score where positive and negative values would indicate that the result was engaged with more and less than expected, respectively, we use the logarithm of COEC.
We also apply additive smoothing to handle items with low action\slash impression counts.
The overall Memboost score is thus
$$\mathrm{MB}(q,r) = \log\frac{\mathrm{actions}(q,r) + \alpha}{\mathrm{Eactions}(q,r) + \alpha}.$$

\subsection{Memboost Scoring}
The Memboost scores are used to adjust the existing scores of pins, and the final results are sorted by this score.
$$\mathrm{MemboostedScore}(q,r) = \mathrm{Score}(q,r) + \gamma\cdot\mathrm{MB}(q,r)$$
Until recently, the Memboost weights $\beta_1, \dots, \beta_4, \gamma$ were hand-tuned through A/B experiments for maximal engagement, but only at a single point in time.
However, this produces an undesirable coupling to the scoring function: experimenting with a new ranker or changing the scoring function could produce larger or smaller initial scores, inadvertently changing the relative magnitude of the Memboost weights.
The hand-tuned weights would then no longer be optimal for the new conditions (system changes, different time period, etc).

To remove this coupling, we now \textit{jointly} retrain the Memboost parameters when changing the model.
We moved to \textit{Memboost as a feature}, where the intermediate Memboost values (clicks, Eclicks, \dots) are fed as features into the machine-learned ranker.

\subsection{Memboost Insertion}
Sometimes, results are known to be good (based on their Memboost scores), but due to upstream changes in the candidate generator and ranking system, the candidate is no longer present in the results. In order to handle these cases, we devised a \textit{Memboost insertion} algorithm which re-inserts the top $n$ results with the highest aggregate Memboost score if they are not already in the incoming result set.

\subsection{Discussion}
Memboost as a whole introduces significant system complexity by adding feedback loops in the system.
It's theoretically capable of corrupting or diluting experiment results: for example, positive results from experiments could be picked up and leaked into the control and production treatments.
It can make it harder to retest past experiments (e.g. new modeling features) after they are launched, because the results from those experiments may already be memorized.

These problems are present in any memorization-based system, but Memboost has such a significant positive impact that we currently accept these implications.

We are currently experimenting with alternatives to Memboost \textit{insertion}, though.
Memboost insertion can slow development velocity, because experiments that harm results may no longer show up as negative A/B results, and the effect of new ranking experiments may be diluted as top results are dominated by Memboost insertion.
Memboost insertion can also indefinitely maintain candidates even if the candidate generator no longer produces them.

A common alternative memorization approach is to incorporate item-id as a ranking feature, such as in~\cite{cheng2016wide}.
However, that requires a large model---linear in the number of items memorized---and consequently a large amount of training data to learn those parameters.
Such large models typically require distributed training techniques.
Instead, Memboost pre-aggregates statistics about the engagement with each result, which allows us to train the main ranking model on a single machine.


\section{Evolution of Ranking}\label{sec:ranking}

Candidate generation and Memboost had already been working for quite some time before ranking was introduced.
We hypothesized the next biggest potential improvement would come from adding a ranking component to our system and applying learning-to-rank techniques~\cite{burges2005learning}\cite{liu2009learningtorank}.
The first learning-to-rank model was an enormous step increase in Related Pins engagement, increasing user propensity to save and click results by over 30\%.

\begin{table}[htbp]
    \small
    \begin{tabulary}{1.0\columnwidth}{L}
        \toprule
        \textbf{Pin Raw Data (aggregated across image signature)} \\
        \midrule
        \textbf{Text annotations}: aggregated from pin descriptions, links, board titles, and other metadata \\ 
        \textbf{Image features}: fc6 and fc8 activations of a deep convolutional neural network \cite{kislyuk2015human} \\
        \textbf{Word embeddings}: aggregated for the text annotations \cite{mao2016wordembeddings} \\
        \textbf{Category vector}: aggregated from user-selected categories for their boards \\
        \textbf{Topic vectors}: computed from the pin-board graph \\
        \textbf{Demographic data}: gender, country, language affinities \\
        
        \midrule
        \textbf{Memboost Data (for query + result pin pair)} \\
        \midrule
        \textbf{Action counts} (clicks, long clicks, saves, closeups) for this result pin. \\
        \textbf{Expected action counts} given the positions in which this result was displayed. \\
        
        \midrule
        \textbf{Offline User Raw Data} \\
        \midrule
        \textbf{Demographic data}: gender, country, language \\
        \textbf{Long- and medium-term activity}: pins saved; annotation, category, and topic vectors \\
        
        \midrule
        \textbf{Real-time User Context} \\
        \midrule
        \textbf{Traffic source}: home feed, search, board, SEO \\
        \textbf{Recent search queries}: tokens and embeddings \\
        \textbf{Recent activity}: pins saved, clicked, closeupped; annotation, category and topic vectors \\
        \bottomrule
    \end{tabulary}
    \caption{Example raw data available to the ranking feature extractor.}
    \label{tab:raw_data}
\end{table}

In our application, the ranker re-orders candidate pins in the context of a particular query $Q$, which comprises the query pin, the viewing user, and user context.
These query components and the candidate pin $c$ each contribute some heterogeneous, structured \textit{raw data}, such as annotations, categories, or recent user activity, shown in Table~\ref{tab:raw_data}.

Our first ranking system only used the pins' raw data.
As we gained additional engineering capacity to build necessary infrastructure, we introduced more data into ranking, such as Memboost and user data.
We also introduced personalized features extracted from users' recent activities, for example users' latest search queries.

We define many \textit{feature extractors} that take this raw data and produce a single feature vector $\phi(Q, c) \in\mathbb{R}^D$.
Some feature extractors directly copy raw data into the feature vector, such as topic and category vectors, while others compute transformations of raw data, such as normalized or re-scaled versions of Memboost data.
Some feature extractors apply one-hot encoding to categorical fields like gender and country.
Finally, some feature extractors compute match scores, such as the category vector cosine similarity between query and candidate pins, or the distance between query and candidate image embeddings.

A ranking model $F : \mathbb{R}^D \to \mathbb{R}$ takes the feature vector and produces a final ranking score.
This ranking model is learned from training data that we describe in the next section.

\subsection{Choices}

We faced three largely orthogonal decisions in building the ranking system: training data collection method, learning objective, and model type.
We first introduce the options we have explored for each choice.

\subsubsection{Training Data Collection}\label{sec:ranking_training_data}

We explored two main sources of training data.

\textbf{Memboost scores as training data.}
Conceptually, the ranker can learn to predict Memboost scores for query-result pairs without enough log data to have a confident Memboost estimate.

\textbf{Individual Related Pins sessions.}
A session is defined as a single user's interactions with Related Pins results from a single query pin.
We can sample these interactions directly as training data.

\subsubsection{Model Objective}\label{sec:ranking_objective}

In \cite{liu2009learningtorank}, learning to rank approaches are broadly categorized into \textit{pointwise}, \textit{pairwise}, and \textit{listwise} approaches.
The main difference between these approaches is whether the loss function considers one, two, or many candidates at a time.
In our work we have explored pointwise and pairwise approaches, compared in Table~\ref{tab:objective_choice}. 

\begin{table}[htbp]
    \small
    \begin{tabulary}{1.0\columnwidth}{Sl L L}
        \toprule
        & \textbf{Classification (Pointwise)}
        & \textbf{Ranking (Pairwise)}
        \\
        \midrule
        \textbf{Data labels}
        & Binary
        & Relative relevance
        \\
        \textbf{Loss function}
        & Considers single candidate
        & Considers difference of scores in candidate pair
        \\
        \bottomrule
    \end{tabulary}
    \caption{Comparison of model objectives.}
    \label{tab:objective_choice}
\end{table}

\subsubsection{Model Formulation}\label{sec:ranking_model}

The precise form of the model determines the model's capacity for describing complex relationships between the features and score.
Table~\ref{tab:model_type_choice} compares two model types that we have used.

\begin{table}[htbp]
    \small
    \begin{tabulary}{1.0\columnwidth}{Sl L L}
        \toprule
        & \textbf{Linear Model}
        & \textbf{Gradient-Boosted Decision Trees}
        \\
        \midrule
        \textbf{Scoring}
        & Linear combination of features
        & Sum of ensemble of piecewise-constant decision trees
        \\
        \textbf{Learning}
        & Learn weight vector by logistic regression or RankSVM
        & Learn decision trees sequentially via gradient boosting
        \\
        \bottomrule
    \end{tabulary}
    \caption{Comparison of model types.}
    \label{tab:model_type_choice}
\end{table}

\subsection{Evolution of our Decisions}

\begin{table*}[htbp]
    \centering
    \small
    \begin{tabulary}{1.0\textwidth}{L L L L L}
        \toprule
        & \textbf{Training Data}
        & \textbf{Label Type}
        & \textbf{Objective}
        & \textbf{Model}
        \\
        \midrule
        V1
        & Memboost
        & Relevance pairs (high MB $>$ low MB ; low MB $>$ random)
        & Pairwise, RankSVM loss 
        & Linear
        \\
        V2
        & Individual sessions
        & Relevance pairs (save $>$ closeup ; ...)
        & Pairwise, RankSVM loss
        & Linear
        \\
        V3
        & Individual sessions
        & Relevance pairs (save $>$ closeup ; ...)
        & Pairwise, RankNet loss
        & GBDT
        \\
        V4
        & Individual sessions
        & Binary (pin saved?)
        & Pointwise, logistic loss
        & GBDT
        \\
        \bottomrule
    \end{tabulary}
    \caption{Evolution of our ranking system. Each row denotes a combination of training data, objective, and model choices that we used.}
    \label{tab:ranking_evolution}
\end{table*}

Table~\ref{tab:ranking_evolution} shows the various combinations of training data, objective, and model that we have explored in Related Pins ranking.

\textbf{Version 1: Memboost training data, relevance pair labels, pairwise loss, and linear RankSVM~\cite{Joachims:2002} model.}
In our first iteration, we chose to use Memboost data because we found it to be the highest-quality signal, by virtue of being an aggregation of millions of users' behavior over a large time period.
We explicitly sampled pairs of pins $(r_1, r_n)$, $(r_n, r_\text{rand})$ for each query, where $r_1, r_n$ are the results with highest and lowest Memboost scores, respectively, for a given query. $r_\text{rand}$ is a randomly generated popular pin from Pinterest, added to stabilize the rankings, as suggested in \cite{Joachims:2002}. We reasoned that pins with low Memboost scores would still be more relevant than purely random pins, because the candidate generator provides some degree of relevance.

When we manually examined pairs from Memboost data, we found that we could guess which pin had a higher Memboost score about 70\% of the time.
This indicated to us that the training data was fairly clean.
(For comparison, pairs sampled from individual user sessions are much noisier; we couldn't discern with any confidence which of two pins was saved by the user.)
Thus, we could use a much smaller corpus and train a model within minutes on a single machine.

\textbf{Version 2: Moved to individual Related Pins sessions.}
We wanted to use user- and context-specific features, but using Memboost data inherently precludes personalization because it is aggregated over many users, losing the association with individual users and session context.
Additionally, we found that only popular content had enough interaction for reliable Memboost data.
These limitations motivated the switch to individual Related Pins sessions.

Each logged session consists of the query pin, viewing user, and recent action context, and a list of result pins. 
Each result also has a corresponding engagement label (one of \textit{impression only}, \textit{closeup}, \textit{click}, \textit{long click}, and \textit{save}). 
For the purposes of training, we trim the logged set of pins, taking each engaged pin as well as two pins immediately preceding it in rank order, under the assumption that the user probably saw the pins immediately preceding the pin they engaged with.
In this iteration, we continued to use a pairwise loss, but with pin relevance pairs defined by a relative ordering of actions: \textit{save} $>$ \textit{long click} $>$ \textit{click} $>$ \textit{closeup} $>$ \textit{impression only}. 

\textbf{Version 3: Moved to a RankNet \cite{burges2005learning} GBDT Model.}
We found that a simple linear model was able to capture a majority of the engagement gain from ranking.
However, linear models have several disadvantages: first, they force the score to depend linearly on each feature.
For the model to express more complex relationships, the engineer must add transformations of these features (bucketizing, percentile, mathematical transformations, and normalization).
Second, linear models cannot make use of features that only depend on the query and not the candidate pin.
For example, if a feature $\phi_k$ represents a feature like "query category = Art", every candidate pin would get the same contribution $w_k \phi_k$ to its score, and the ranking would not be impacted.
The features specific to the query must be manually crossed with candidate pin features, such as adding a feature to represent "query pin category + candidate category".
It is time consuming to engineer these feature crosses.

To avoid these downsides, we moved to \textit{gradient-boosted decision trees} (GBDT).
Besides allowing non-linear response to individual features, decision trees also inherently consider interactions between features, corresponding to the depth of the tree.
For example, it becomes possible to encode reasoning such as "if the query pin's category is Art, visual similarity should be a stronger signal of relevance."
By automatically learning feature interactions, we eliminate the need to perform manual feature crosses, speeding up development.

\textbf{Version 4: Moved to pointwise classification loss, binary labels, and logistic GBDT model.}
Although we initially opted for pairwise learning, we have since attained good results with pointwise learning as well.
Since our primary target metric in online experiments is the propensity of users to save result pins, using training examples which also include closeups and clicks seemed counterproductive since these actions may not reflect save propensity.
We found that giving examples simple binary labels ("saved" or "not saved") and reweighting positive examples to combat class imbalance proved effective at increasing save propensity.
We may still experiment with pairwise ranking losses in the future with different pair sampling strategies.

\subsection{Previous-Model Bias}

\begin{figure}[htbp]
    \centering
    \begin{subfigure}[t]{0.8\columnwidth}
        \includegraphics[width=\columnwidth]{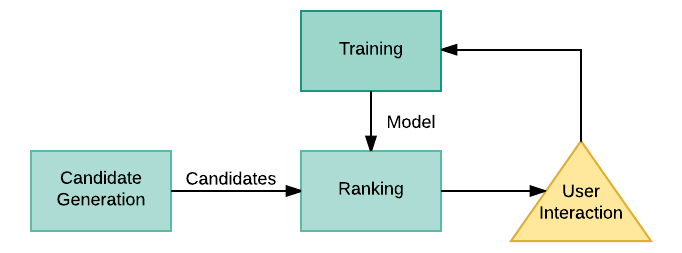}
        \caption{Direct feedback loop}
        \label{fig:loop}
    \end{subfigure}
    \begin{subfigure}[t]{0.8\columnwidth}
        \includegraphics[width=\columnwidth]{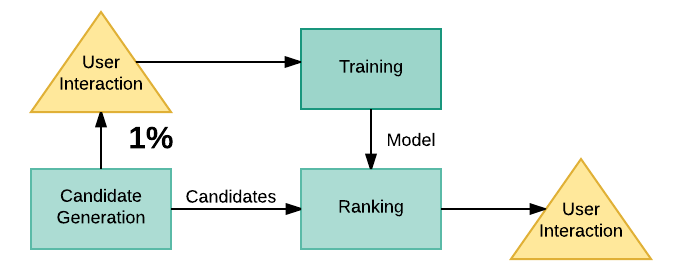}
        \caption{No feedback loop}
        \label{fig:no_loop}
    \end{subfigure}
    \caption{We removed a feedback loop by collecting randomized training data.}
    \label{fig:loop_vs_no_loop}
\end{figure}

During our efforts to improve ranking, we experienced a major challenge.
Because engagement logs are used for training, we introduced a \textit{direct feedback loop}, as described in \cite{sculleyhidden}: the model that is currently deployed dramatically impacts the training examples produced for future models.
We directly observed the negative impact of this feedback loop.
When we trained the first ranking model, the logs reflected user's engagement with results ranked only by the candidate generator. The learned model was applied to rank these same candidates.
Over the following months, the training data only reflected engagement with pins that were \textit{highly ranked by the existing model} (Figure~\ref{fig:loop}).
When we tried to train a model with the same features but with the latest engagement data, we were unable to beat the already-deployed model.
We hypothesized that the feedback loop posed a problem since the distribution of training pins no longer matched the distribution of pins ranked at serving time.

To alleviate this "previous-model" bias in the training data, we allocate a small percentage of traffic for "unbiased data collection": for these requests, we show a random sample from all our candidate sources, randomly ordered without ranking.
This isolates the training data from being influenced by the previous model (Figure~\ref{fig:no_loop}).
Although the unranked results are lower quality, they provide valuable data for training new ranking models.
To avoid degrading any particular user's experience too much, each user is served unranked pins on only a small random subset of queries.
Although the volume of training data becomes limited to this small percentage of overall traffic, the resulting models perform better than models trained with biased data.

\subsection{Success Metrics}\label{sec:ranking_offline_eval}
One important step in being able to explore these different options is being able to iterate quickly.
The gold standard for testing changes is online A/B experimentation, where we evaluate ranking primarily by their impact on save propensity.

All changes undergo online experiments, but this process takes days or weeks to collect data.
We find it helpful to test changes immediately via offline evaluation to approximate different models' performance.
In this process, we reuse much of our training data generator to sample individual Related Pins sessions, but choose a distinct range of dates that follows the training date range, and a slightly different sampling strategy.
For each session we rescore the pins that the user actually saw, using the models under test, then measure the agreement between the scores and the logged user behavior.

We have experimented with various measures, including normalized discounted cumulative gain (NDCG), area under the precision-recall curve (PR AUC) with various interpolation methods, and precision vs.\ position AUC.

To determine how well these offline evaluation metrics predict live A/B experiment impact, we examined the results for several of our past ranking model changes.
We examined the directionality as well as the magnitude of the difference predicted by offline evaluation, and compared it to the actual experiment results.
We found that PR AUC metrics are extremely predictive of closeups and clickthroughs in A/B experiments, but we had difficulty predicting the save behavior using offline evaluation.
For now, we use offline metrics as a sanity check and rough estimation of potential impact.

\subsection{Serving Infrastructure}
Related Pins serves many tens of thousands of queries per second at peak loads.
To handle this scale, we leverage several existing Pinterest systems.

Our first version of pin ranking was pre-computed in offline map-reduce jobs and served from Terrapin \cite{sharma2015terrapin}, an immutable key-value lookup service.
This required a massive map-reduce job to join raw data for every query and candidate, compute features and score the items.

Due to limitations of the cluster we could only rank a few million queries at a time, yielding 50\% coverage of user queries.
We scaled up by running the reranking job on different segments at a time and combining the results, but this approach was inherently unable to give full coverage in a reasonable amount of time.
Offline ranking also significantly slowed development velocity: each experiment that changed the model (feature development, training data changes) required reranking all queries offline, a time-consuming process.

Thus, we moved to an online ranking serving system.
The pin raw data is stored on a sharded key-value store called RealPin \cite{liu2015realpin}, keyed by image signature.
To perform ranking, we assemble a request with the list of candidate pins and other raw data that will be needed to compute the features and score: query pin raw data (retrieved from RealPin), offline user raw data (from Terrapin), recent user activity (from a service called UserContextService), and Memboost data (from Terrapin).
The RealPin root server replicates the request to the leaves, routing the appropriate subset of the candidates to each leaf server.
The leaves locally retrieve pin raw data and invoke our custom feature extractor and scorer.
The leaves send the top candidates and scores to the root node, which gathers and returns the overall top candidates.

We chose this serving architecture to increase data locality.
The Hadoop-based system suffered from having to transfer a huge amount of pin raw data for each query.
We have also seen that other online pin scoring systems were network-bound due to transfer of pin raw data.
By pushing computation down to the nodes that store the candidates' pin raw data, the bulk of the data transfer can be avoided.


\section{Challenges}\label{sec:challenges}

\subsection{Changing Anything Changes Everything}
According to \cite{sculleyhidden} machine-learning systems inherently tangle signals; inputs are never really independent, resulting in the Changing Anything Changes Everything (CACE) principle: one component of a system can be highly optimized for the existing state of the system.
Improving another component may actually result in worse overall performance. 
This is a \textit{system-level local optimum} that can make further progress difficult. 
Our general solution is to jointly train/automate as much of the system as possible for each experiment. 

We present several examples where this particular challenge has appeared within our simple recommendation system and our mitigations.

\textbf{Example 1: }
Many parameters are used in various stages in our recommender pipeline. 
Recall that we used hand-tuned weights for Memboost optimized with time- and labor-intensive A/B experiments; these became quickly outdated as other parts of the system changed.
Jointly training Memboost weights avoids this problem.
Similarly, the ranking learners have hyperparameters that must be tuned.
To avoid other changes resulting in hyperparameters becoming suboptimal, we implemented a parallelized system for automated hyperparameter tuning.
As a result, we can now optimize hyperparameters each time we change the model.

\textbf{Example 2: }
"Improvements" to the raw data can harm our results since our downstream model is trained on the old definition of the feature.
Even if a bug is fixed, for example in the computation of pin category vectors, our existing model would depend on the flawed definition of the feature, so the fix may negatively impact our system.
This is especially problematic if the change originates from another team.
In \cite{sculleyhidden}, this is termed an \textit{unstable data dependency}.

Currently we must manually retrain our model with updated raw data, and deploy the new model and raw data into production at the same time.
In addition to being time-consuming, this solution is less-than-ideal since it requires us to know of the upstream change.
Ideally, we would automate the continual retraining of our model, which would incorporate any change in upstream data.

\textbf{Example 3: } 
Finally, we have experienced complex interdependencies between candidate generation and ranking.
The model becomes attuned to the idiosyncrasies of the training data.
For example, changing or introducing a candidate generator can cause the ranker to become worse, since the training data distribution will no longer match the distribution of data ranked at serving time; an issue we saw in training data collection.
If introducing a candidate generator does not result in a performance improvement, how does one determine if this was because the candidate generator is poor or because the ranker was not trained on candidates from that candidate generator?
Our current solution is to insert the new candidates into the training data collection for some time before running an experiment with a newly trained model.

This problem highlights the need to simplify the system as much as possible, because the number of possible unintended interactions increases rapidly with the number of system components.

\subsection{Content Activation}
There is a large amount of content without much engagement, but that could be potentially relevant and high quality.
Millions of new images are uploaded to Pinterest each day.
Additionally, there is a large pool of "dark" content that is high quality but rarely surfaced.
Balancing fresh and dark content with well-established, high-quality content represents the classic explore vs.\ exploit problem, which is an open problem at Pinterest. We dive deeper into how we solved for content activation for the particular case of localization.

Since Pinterest is an increasingly international product, we made a special effort to ensure that international users saw content in their own language. 
Making Related Pins local was difficult for two main reasons. 
First, there were not many local candidates in the first place, since the content did not appear on many boards.
Second, there was much less historical engagement with local content, which caused our models to rank it lower, even if we did have local candidates.

\begin{figure}[htbp]
    \centering
    \includegraphics[width=\columnwidth]{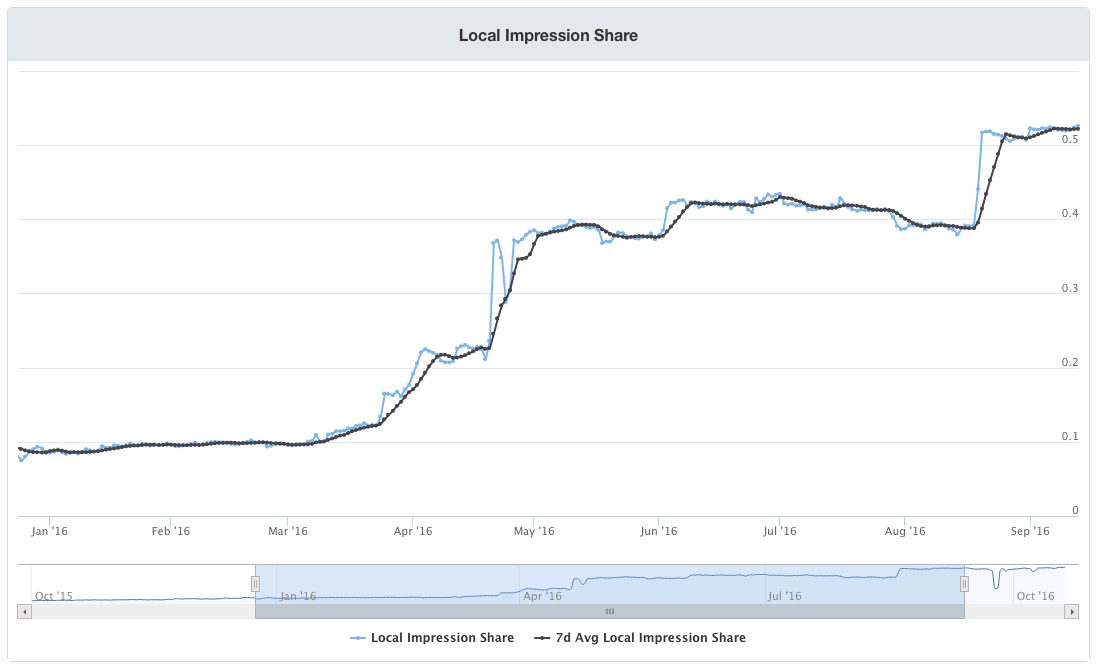}
    \caption{Proportion of Related Pins impressions that are local (in the user's native language) for our major international countries.}
    \label{fig:local}
\end{figure}

We tried several approaches to solve these issues, and were eventually able to increase the fraction of local results from 9\% local to 54\% local as shown in Figure~\ref{fig:local}.

\paragraph{Local pin swap}
For each related pin we generate, we check if there is a local alternative with the same image. If so, then we swap the result pin with a pin in the viewer's language.
As a result we increased local impressions and local pins saved, without changing the relevance of the results.

\paragraph{Local pin boost}
When result pins existed that were in the viewer's language, we attempted to artificially promote them to a higher position in the result set. This did not prove to be particularly effective, because the candidates at the time did not contain many local pins, resulting in low coverage for this solution.

\paragraph{Localizing candidate sets}
We modified our board-based candidate generation method to filter on language before sampling pins, producing a segmented corpus of pins for each language, described in Section~\ref{sec:candidates/segmented}.
Furthermore, as a way of increasing exposure to these local pins, we chose to blend local candidates into the results at various ratios.


\section{Conclusion}
Recommender systems literature showcase many impressive state-of-the-art systems, which usually reflect years of iteration.
Yet it's common in industry to prioritize the simplest, highest-leverage solutions.
We offer an inside look at the process by which a recommender system can be conceived and developed by a small engineering team, with limited computational resources and noisy Web-scale data.
We illustrate how organic system growth leads to complexity, and how the CACE principle is indeed pervasive in real-world recommender systems.
We identify unique interdependencies that made it hard to reason about changes in Related Pins, and propose to mitigate these issues by automated joint training of system components.
We tackle the challenge of activating unconnected content, both with regards to the "cold-start" problem and the "rich get richer" problem that is prevalent among systems that exploit engagement signals.
In these solutions, it was important to diversify content, because engagement is not always correlated to relevance.
Finally, making more of the system real-time, both in candidate generation and ranking, significantly increased velocity of experimentation and improved responsiveness of the results.


\section{Acknowledgments}

We thank Dmitry Chechik, Hui Xu, and Yuchen Liu, for developing the original version of board-based Related Pins;
Bo Liu, Shu Zhang, and Jian Fang, for serving infrastructure components;
Vanja Josifovski, Xin Liu, and Mukund Narasimhan for sharing machine learning insight and expertise;
Jure Leskovec, Pong Eksombatchai, and Mark Ulrich for the Pixie service;
Sarah Tavel, Maura Lynch, and Frances Haugen for product management support;
Patrick Phelps for data science support.
Thanks also to Yuchen Liu for Figure~\ref{fig:occur_examples}.


\balance
\bibliographystyle{abbrv}
\bibliography{citation}

\end{document}